\documentclass[12pt]{article}
\usepackage{cite,epsfig,amssymb,amsmath,graphicx,color}
\newcommand{\nn}{\nonumber}

\begin{document}

\vspace{9mm}

\begin{center}
{{{\Large \bf  Abelian Gauge Invariance of the WZ-type Coupling in ABJM Theory}}
\\[17mm]
Dongmin Jang$^{1}$,~~Yoonbai Kim$^{1}$,~~O-Kab Kwon$^{3,4}$,~~D.~D. Tolla$^{2}$
\\[3mm]
{\it $^{1}$Department of Physics,~BK21 Physics Research Division,~Institute of Basic Science, Sungkyunkwan University, Suwon 440-746, Korea
\\
$^{2}$International School for Advanced Studies (SISSA), Via Bonomea 265, 34136 Trieste, Italy}
\\[2mm]
$^3${\sl Department of Physics, Kyungpook National University, Taegu 702-701, Korea}
\\[2mm]
{\it $^{4}$Department of Physics, Ewha Womans University, Seoul 120-750, Korea}\\[2mm]
{\it dongmin@skku.edu,~yoonbai@skku.edu,~okabkwon@ewha.ac.kr,~ddtolla@skku.edu} }

\end{center}
\vspace{20mm}

\begin{abstract}
We construct the interaction terms between the worldvolume fields of multiple M2-branes and 3-form gauge field of 11-dimensional supergravity, in the context of ABJM theory.
The obtained Wess-Zumino-type coupling is simultaneously invariant under the U$_{\textrm{L}}(N)\times$U$_{\textrm{R}}(N)$ non-Abelian gauge transformation of the ABJM theory and {\it the Abelian gauge transformation} of the 3-form field in 11-dimensional supergravity.

\end{abstract}

\newpage
\tableofcontents

\section{Introduction}

In type IIA and IIB string theories, the RR form fields in 10-dimensional supergravities are coupled to the D-branes through Wess-Zumino(WZ)-type action \cite{Li:1995pq,Douglas:1995bn,Myers:1999ps}.
In the effective field theory of multiple D$p$-branes, the WZ-type action includes the couplings to higher rank RR form fields, which are usually referred to as the Myers couplings \cite{Myers:1999ps}.
Like the WZ-type couplings of D-branes in string theory, WZ-type couplings of multiple M2-branes can be constructed \cite{Li:2008ez,Ganjali:2009kt,Kim:2009nc,Lambert:2009qw,Sasaki:2009ij,Kim:2010hj,Allen:2011pm} in the context of the effective field theories, for instance, the Bagger-Lambert-Gustavsson theory \cite{Bagger:2007jr} and the Aharony-Bergman-Jafferis-Maldacena(ABJM) theory \cite{Aharony:2008ug}.
These WZ-type couplings describe the couplings between M2-branes and 3- and 6-form gauge fields in 11-dimensional supergravity.

In \cite{Kim:2010hj} the invariance under the non-Abelian gauge symmetry, U$_{\textrm{L}}(N)\times$U$_{\textrm{R}}(N)$ of the original ABJM theory \cite{Aharony:2008ug}, was utilized to determine the WZ-type couplings on the M2-brane worldvolume.
The results were extended to include non-linear terms of the form fields \cite{Allen:2011pm}.
The proposed WZ-type action in \cite{Kim:2010hj} was put to some tests and proven to be consistent.
First, in the particular case of $N=1$, it nicely reproduces the well-known coupling of the 3-form gauge field to the worldvolume fields of a single M2-brane \cite{Bergshoeff:1987cm}.
Second, under the circle compactification, the action gives the correct Myers coupling of the RR form fields to the worldvolume fields of D2-branes in type IIA string theory \cite{Myers:1999ps}.
Third, in the particular case of a 6-form gauge field with constant 7-form field strength, the proposed WZ-type action in \cite{Lambert:2009qw,Kim:2010hj} reproduces the  full supersymmetry-preserving quadratic mass-deformation of the ABJM theory \cite{Hosomichi:2008jb,Gomis:2008vc}.
Less supersymmetric cases of ${\cal{N}}=2$ and ${\cal{N}}=4$ in ABJM theory have also been investigated in \cite{Kim:2011qv,Kim:2012gz}.
The aforementioned tests support the correctness of the proposed WZ-type coupling to a reasonable extent, however, there remains one more important test to be passed, i.e. the invariance under the Abelian gauge transformation of the form fields in 11-dimensional supergravity.
It is the main goal of this paper to conduct this test.

The 11-dimensional supergravity action is invariant under  
{\it the Abelian gauge transformation} of the form fields,
\begin{align}\label{C36}
C_{r}
\rightarrow
C_{r}
+
d\Lambda_{r-1}
,
\end{align}
where $r=3,6$.
Therefore, the WZ-type couplings on the worldvolume of M2-branes should also satisfy the invariance under \eqref{C36}.
For the Myers couplings of RR form fields, this issue was clarified in \cite{Ciocarlie:2001qv,Adam:2003uq,Adam:2005wx}.
In this paper, we consider the  WZ-type couplings for the 3-form gauge field in the viewpoint of such {\it Abelian gauge invariance.}
We show that the WZ-type couplings in \cite{Kim:2010hj} is invariant under the Abelian gauge transformation \eqref{C36} only when the field strengths, $F_{\mu\nu}$ and $\hat{F}_{\mu\nu}$ of the non-Abelian gauge fields of the U$_{\textrm{L}}(N)\times$U$_{\textrm{R}}(N)$ gauge symmetry, are vanishing.
In the case of non-vanishing non-Abelian field strengths, we show that the coupling needs a modification by a piece involving those field strengths, in order to be invariant under the Abelian gauge transformation.
We find an exact form of the modification and propose a simple form of the 3-form field couplings, which resemble the case of the Myers couplings \cite{Myers:1999ps} in string theory.

This paper is organized as follows.
In section 2, we test the Abelian gauge invariance of the 3-form field couplings in all orders of the expansion parameter with vanishing non-Abelian gauge field strengths, proposed in \cite{Kim:2010hj}.
In section 3, we propose a simple form of the 3-form field couplings with non-vanishing non-Abelian gauge field strengths and test the proposal is invariant under the Abelian gauge transformation.
In section 4 we draw our conclusion.

\section{Abelian Gauge Invariance: $F_{\mu\nu}=\hat{F}_{\mu\nu}=0$ Case}

In the ABJM theory of multiple M2-branes, the bosonic sector of the M2-brane worldvolume fields contains two non-Abelian gauge fields, $A_{\mu}$ and $\hat{A}_{\mu}$, and four complex scalar fields, $Y^{A}$, ($A=1,2,3,4$).
The WZ-type couplings were constructed by using four covariant building blocks and their complex conjugates \cite{Kim:2010hj}.
These building blocks are the 3-form gauge field $C_{3}$, the 6-form gauge field $C_{6}$, both of which are functionals of the complex scalar fields, the covariant derivatives of the complex scalar fields $D_{\mu}Y^{A}=\partial_{\mu}Y^{A}+iA_{\mu}Y^{A}-iY^{A}\hat{A}_{\mu}$, and the anti-symmetrized cubic product of the complex scalar fields, $\beta^{AB}_{~C}\equiv\frac{1}{2}(Y^{A}Y^{\dagger}_{C}Y^{B}-Y^{B}Y^{\dagger}_{C}Y^{A})$.
The manifestly covariant objects, but missing from this list, are the non-Abelian gauge field strengths, $F_{\mu\nu}$ and $\hat{F}_{\mu\nu}$.
In \cite{Kim:2010hj}, the WZ-type couplings are constructed under the assumption that these gauge field strengths are vanishing, which means that the corresponding non-Abelian gauge fields were in pure gauge.
In this section, we reconsider the 3-form WZ-type couplings proposed in \cite{Kim:2010hj} and show that those are invariant in all orders of the expansion parameter under the Abelian gauge transformation \eqref{C36}, when the non-Abelian gauge fields, $A_{\mu}$ and $\hat{A}_{\mu}$, are in pure gauge, i.e. $F_{\mu\nu}=\hat{F}_{\mu\nu}=0$.

\subsection{Definitions}

In order to show the Abelian gauge invariance for the 3-form WZ-type couplings, let us consider such type of coupling for a generic $p$-form gauge field, which does naturally couple to a $(p-1)$-brane.
Eventually, we specialize the results to the $p=3$ case.
The specific form of WZ-type couplings is given by
\begin{align}\label{Sp}
\tilde{S}_{p}
=
\mu_{p-1}\int_{p}
\{\textrm{Tr}_{S}\}
P[C_{(p)}]
=
\frac{\mu_{p-1}}{2}\int_{p}d^{p}x
\{\textrm{Tr}_{S}\}
\frac{1}{p!}\epsilon^{\mu_{1}\cdots\mu_{p}}
\left(
P[C_{(p)}]_{[\mu_{1}\cdots\mu_{p}]}
+(\textrm{c.c.})
\right)
,
\end{align}
where $\mu_{p-1}$ represents the tension of $(p-1)$-brane, $P[\cdots]$ is a non-Abelian pullback (see below or \cite{Kim:2010hj}), $\{\textrm{Tr}_{S}\}$ denotes the sum over all possible ways that the gauge indices can be contracted to form a single trace product divided by the number of independent terms at a given order in the expansion parameter $\lambda$.
More precisely, $\{\textrm{Tr}_{S}\}=\{\textrm{Tr}\}/n_{\textrm{terms}}$, where $\{\textrm{Tr}\}$ is defined in \cite{Kim:2010hj} and $n_{\textrm{terms}}$ is the number of independent terms at a given order in $\lambda$.
Generalizing the definitions given in \cite{Kim:2010hj}, the non-Abelian pullback of the $p$-form gauge field is given by
\begin{align}\label{CpPB}
P[C_{(p)}]_{\mu_{1}\cdots\mu_{p}}
=&
C_{A_{1}\cdots A_{m}\bar{B}_{1}\cdots\bar{B}_{n}}
\left(
\delta^{A_{1}}_{\mu_{1}}I_{N}
+\lambda D_{\mu_{1}}Y^{A_{1}}
\right)
\cdots
\left(
\delta^{A_{m}}_{\mu_{m}}I_{N}
+\lambda D_{\mu_{m}}Y^{A_{m}}
\right)
\nn\\
&
\left(
\delta^{\bar{B}_{1}}_{\mu_{m+1}}I_{N}
+\lambda D_{\mu_{m+1}}Y^{\dagger}_{B_{1}}
\right)
\cdots
\left(
\delta^{\bar{B}_{n}}_{\mu_{m+n}}I_{N}
+\lambda D_{\mu_{m+n}}Y^{\dagger}_{B_{n}}
\right)
\nn\\
=&\left(\begin{matrix}p \\ l\end{matrix}\right)
\left(\begin{matrix}p-l \\ k\end{matrix}\right)
C_{A_{1}\cdots A_{l}\bar{B}_{1}\cdots\bar{B}_{k}[\mu_{l+k+1}\cdots\mu_{p}}D_{\mu_{1}}Y^{A_{1}}\cdots D_{\mu_{l}}Y^{A_{l}}D_{\mu_{l+1}}Y^{\dagger}_{B_{1}}\cdots D_{\mu_{l+k}]}Y^{\dagger}_{B_{k}}
,
\end{align}
where $\left(\begin{matrix}m \\ n\end{matrix}\right)
=\frac{m!}{(m-n)!n!},$ $\lambda=2\pi l^{3/2}_{\textrm{P}}$ ($l_{\textrm{P}}$ is the Planck length), and $I_{N}$ is the $N\times N$ unit matrix.
Using such definition of the pullback for the $p=3$ case, the WZ-type coupling in \eqref{Sp} gives
\begin{align}\label{3action}
\tilde{S}_{3}
=
\mu_{2}\int d^{3}x\frac{\epsilon^{\mu\nu\rho}}{3!}
&
\{\textrm{Tr}_{S}\}
\left[
\frac{1}{2}C_{\mu\nu\rho}
+3\lambda C_{\mu\nu A}D_{\rho}Y^{A}
+3\lambda^{2}
\left(
C_{\mu AB}D_{\nu}Y^{A}D_{\rho}Y^{B}
+C_{\mu A\bar{B}}D_{\nu}Y^{A}D_{\rho}Y^{\dagger}_{B}
\right)
\right.
\nn\\
&
\left.
+\lambda^{3}
\left(
C_{ABC}D_{\mu}Y^{A}D_{\nu}Y^{B}D_{\rho}Y^{C}
+3C_{AB\bar{C}}D_{\mu}Y^{A}D_{\nu}Y^{B}D_{\rho}Y^{\dagger}_{C}
\right)
+(\textrm{c.c.})
\right]
,
\end{align}
where $\mu_{2}$ is the tension of M2-brane.
This form of the WZ-type couplings was proposed in \cite{Kim:2010hj}.

Note that the background form fields are 
functions of the transverse coordinates in general, so they become functionals 
of the transverse scalar fields, $Y$ and $Y^\dagger$. 
The dependence of the 3-form gauge field on the complex scalar fields is expressed by means of a generalized Taylor expansion,
\begin{align}\label{Taylor}
C(Y,Y^{\dagger})
=
\sum_{r,s}\frac{\lambda^{r+s}}{r!s!}Y^{A_{1}}\cdots Y^{A_{r}}Y^{\dagger}_{B_{1}}\cdots Y^{\dagger}_{B_{s}}\partial_{A_{1}}\cdots\partial_{A_{r}}\partial_{\bar{B}_{1}}\cdots\partial_{\bar{B}_{s}}C^{0}
,
\end{align}
where the superscript `$0$' means that the corresponding field has no dependence on the complex scalar fields and we omit the indices on $3$-form gauge field, and $\partial_{A}\equiv\frac{\partial}{\partial (\lambda Y^{A})}$, $\partial_{\bar{B}}=\bar{\partial}_{B}\equiv\frac{\partial}{\partial(\lambda Y^{\dagger}_{B})}$, $(\partial\bar{\partial}\cdots)C^{0}\equiv(\partial\bar{\partial}\cdots)C(Y,Y^{\dagger})|_{Y=Y^{\dagger}=0}$.
Keeping \eqref{Taylor} in mind, for the U$_{\textrm{L}}(N)\times$U$_{\textrm{R}}(N)$ gauge invariance, each term of the WZ-type couplings in \eqref{3action} should contain equal numbers of bifundamental fields ($Y$, $DY$) and anti-bifundamental fields ($Y^{\dagger}$, $DY^{\dagger}$).
In addition, the gauge indices must be contracted appropriately to form a single trace coupling.

Using \eqref{CpPB} and inserting the expanded $p$-form gauge field \eqref{Taylor} into the action \eqref{Sp}, we obtain the WZ-type couplings for the $p$-form gauge field in terms of the expansion parameter $\lambda$,
\begin{align}\label{Sp2}
\tilde{S}_{p}
=
\frac{\mu_{p-1}}{2}\int_{p}d^{p}x\frac{1}{p!}
\left(
\epsilon^{\mu^{(p)}}\sum_{r,s}\sum_{l+k=0}^{p}\lambda^{2q}b^{l,r}_{k,s}
+(\textrm{c.c.})
\right)
,
\end{align}
where $l$, $k$ are the numbers of $DY$, $DY^{\dagger}$ from the pullback in \eqref{CpPB} and $r$, $s$ are the numbers of $Y$, $Y^{\dagger}$ from the Taylor expansion in \eqref{Taylor}, and
\begin{align}\label{bkslr}
b^{l,r}_{k,s}
=
U^{l,r}_{k,s}
\{\textrm{Tr}_{S}\}
\left[
(D_{\mu}Y^{A})^{(l)}(D_{\nu}Y^{\dagger}_{B})^{(k)}Y^{C(r)}Y^{\dagger(s)}_{D}
\right]
\end{align}
with
\begin{align}\label{Ukslr}
U^{l,r}_{k,s}
=
\frac{1}{r!s!}\left(\begin{matrix}p \\ l\end{matrix}\right)
\left(\begin{matrix}p-l \\ k\end{matrix}\right)
\partial_{C^{(r)}}\partial_{\bar{D}^{(s)}}C^{0}_{A^{(l)}
\bar{B}^{(k)}\mu^{(p-l-k)}}.
\end{align}
In order to avoid a cluttering of our expressions, we have introduced the following compact notation for our indexing
\begin{align}\label{indexing}
&
\mu^{(p)}
\equiv
\mu_{1}\cdots\mu_{p}
,
\nn\\
&
\partial_{A^{(r)}}
\equiv
\partial_{A_{1}}\cdots\partial_{A_{r}}
,
\nn\\
&
(D_{\mu}Y^{A})^{(l)}
\equiv
D_{\mu_{1}}Y^{A_{1}}\cdots D_{\mu_{l}}Y^{A_{l}}
,
\nn\\
&
Y^{C(r)}
\equiv
Y^{C_{1}}\cdots Y^{C_{r}}
,
\,\,\,
\textrm{etc}
.
\end{align}
We also use indices ($\mu$, $A$), ($\nu$, $B$) only with $DY$, $DY^{\dagger}$, while the indices ($C$, $D$) are used only with ($Y$, $Y^{\dagger}$).

The WZ-type couplings in \eqref{Sp2} are originated from the $p$-from gauge field $C_{p}$.
Therefore, the number of covariant derivatives involved must be less than or equal to $p$, which means
\begin{align}\label{cnsts1}
0
\le
l
+k
\le
p
.
\end{align}
The coupling should also be invariant under the non-Abelian gauge symmetry, U$_{\textrm{L}}(N)\times$U$_{\textrm{R}}(N)$, of which realization requires that the number of involved bifundamental and anti-bifundamental fields must be the same, i.e.
\begin{align}\label{cnsts}
l
+r
=
k
+s
=
q
,
\end{align}
where $q$ is the total number of $Y$ and $DY$ (or $Y^{\dagger}$ and $DY^{\dagger}$) in a given term of the WZ-type coupling.
Using the constraints in \eqref{cnsts1} and \eqref{cnsts}, we rewrite the WZ-type coupling \eqref{Sp2} as
\begin{align}\label{Sp3}
\tilde{S}_{p}
=
\frac{\mu_{p-1}}{2}\int_{p}d^{p}x\frac{1}{p!}
\left(
\epsilon^{\mu^{(p)}}\sum_{q=0}^{\infty}\sum_{m=0}^{p}\sum_{k=0}^{m}\lambda^{2q}\,b^{m-k,q-m+k}_{k,q-k}
+(\textrm{c.c.})
\right)
.
\end{align}

\subsection{Abelian gauge invariance}

To prove the Abelian gauge invariance \eqref{C36} for the WZ-type couplings in \eqref{Sp2}, we repeatedly integrate by parts the quantity $b^{l,r}_{k,s}$.
Then the expression \eqref{bkslr} can be written completely in terms of a $(p+1)$-form field strengths,
\begin{align}\label{Fback}
F^{0}_{\mu\nu^{(i)}A^{(j)}\bar{B}^{(k)}}
=
(p+1)\partial_{[\mu}C^{0}_{\nu^{(i)}A^{(j)}\bar{B}^{(k)}]}
,
\qquad
i
+j
+k
=
p
,
\end{align}
where we used the compact indexing notation defined in \eqref{indexing}.
Once this procedure is achieved, the resulting expression is manifestly gauge invariant because of the Abelian gauge invariance of the $(p+1)$-form field strengths.

First, let us consider the case $l\ne0$.
Integrating by parts, $b^{l,r}_{k,s}$ can be written as
\begin{align}\label{bkslr2}
b^{l,r}_{k,s}
=&
U^{l,r}_{k,s}
\{\textrm{Tr}_{S}\}
\left[
(D_{\mu}Y^{A})^{(l-1)}D_{\mu'}Y^{A'}(D_{\nu}Y^{\dagger}_{B})^{(k)}Y^{C(r)}Y^{\dagger(s)}_{D}
\right]
\nn\\
=&
G^{l,r}_{k,s}
-E^{l,r}_{k,s}
-rA^{l,r}_{k,s}
-sB^{l,r}_{k,s}
,
\end{align}
where we omit the total derivative term and
\begin{align}
G^{l,r}_{k,s}
=&
-(l-1)U^{l,r}_{k,s}
\{\textrm{Tr}_{S}\}
\left[
(D_{\mu}Y^{A})^{(l-2)}D_{\mu'}D_{\mu''}Y^{A''}Y^{A'}(D_{\nu}Y^{\dagger}_{B})^{(k)}Y^{C(r)}Y^{\dagger(s)}_{D}
\right]
\nn\\
&
-kU^{l,r}_{k,s}
\{\textrm{Tr}_{S}\}
\left[
(D_{\mu}Y^{A})^{(l-1)}Y^{A'}(D_{\nu}Y^{\dagger}_{B})^{(k-1)}D_{\mu'}D_{\nu'}Y^{\dagger}_{B'}Y^{C(r)}Y^{\dagger(s)}_{D}
\right]
,
\nn\\
E^{l,r}_{k,s}
=&
\left(
\partial_{\mu'}U^{l,r}_{k,s}
\right)
\{\textrm{Tr}_{S}\}
\left[
(D_{\mu}Y^{A})^{(l-1)}Y^{A'}(D_{\nu}Y^{\dagger}_{B})^{(k)}Y^{C(r)}Y^{\dagger(s)}_{D}
\right]
,
\nn\\
A^{l,r}_{k,s}
=&
U^{l,r}_{k,s}
\{\textrm{Tr}_{S}\}
\left[
(D_{\mu}Y^{A})^{(l-1)}Y^{A'}(D_{\nu}Y^{\dagger}_{B})^{(k)}Y^{C(r-1)}D_{\mu'}Y^{C'}Y^{\dagger(s)}_{D}
\right]
,
\nn\\
B^{l,r}_{k,s}
=&
U^{l,r}_{k,s}
\{\textrm{Tr}_{S}\}
\left[
(D_{\mu}Y^{A})^{(l-1)}Y^{A'}(D_{\nu}Y^{\dagger}_{B})^{(k)}Y^{C(r)}Y^{\dagger(s-1)}_{D}D_{\mu'}Y^{\dagger}_{D'}
\right]
.
\end{align}
Here worldvolume indices are anti-symmetrized but are kept implicit.
Therefore, the presence of the two covariant derivatives acting on a single object implies that such terms contain the non-Abelian gauge field strengths due to the relation,
\begin{align}
\left[
D_{\mu}
,D_{\nu}
\right]
Y^{A}
=
iF_{\mu\nu}Y^{A}
-iY^{A}\hat{F}_{\mu\nu}
.
\end{align}
For this reason, we see that $G^{l,r}_{k,s}$ terms in \eqref{bkslr2} are vanishing in the case of $F_{\mu\nu}=\hat{F}_{\mu\nu}=0$.
As a result, if the WZ-type coupling in \eqref{Sp2} can be rewritten in terms of $F^{0}_{(p+1)}$ and $G^{l,r}_{k,s}$, then that is enough to prove the invariance of \eqref{Sp2} under the Abelian gauge transformation when $F_{\mu\nu}=\hat{F}_{\mu\nu}=0$.

The expression of $U^{l,r}_{k,s}$ in \eqref{Ukslr} contains $\partial C^{0}$.
In order to convert such terms to a $(p+1)$-form field strength, we need to totally anti-symmetrize the indices on $\partial C^{0}$ as follows
\begin{align}\label{antisym}
\partial_{\alpha}C^{0}_{\beta_{1}\cdots\beta_{p}}
=
(p+1)\partial_{[\alpha}C^{0}_{\beta_{1}\cdots\beta_{p}]}
+\partial_{\beta_{1}}C^{0}_{\alpha\beta_{2}\cdots\beta_{p}}
+\cdots.
\end{align}
Here the first term of \eqref{antisym} is a component of the $(p+1)$-form field strength and so it is invariant under the  gauge transformation \eqref{C36}.
The reaming terms are not gauge invariant, therefore, there should be a complete cancellation of such terms in order to guarantee the gauge invariance.
This is what we are going to show next.

Using the anti-symmetrization in \eqref{antisym}, one can rewrite $A^{l,r}_{k,s}$ and $B^{l,r}_{k,s}$ for $l\ne0$ as
\begin{align}\label{ABkslr}
A^{l,r}_{k,s}
&=
\frac{1}{l}b^{l,r}_{k,s}
+\frac{p+1}{l}{F_{A}}^{l,r}_{k,s}
+\frac{(l+1)(s+1)}{lr}B^{l+1,r-1}_{k-1,s+1}
-\frac{l+1}{lr}E^{l+1,r-1}_{k,s}
,
\nn\\
B^{l,r}_{k,s}
&=
\frac{r+1}{ls}b^{l-1,r+1}_{k+1,s-1}
+\frac{p+1}{k+1}{F_{B}}^{l,r}_{k,s}
-\frac{(l-1)(r+1)}{ls}A^{l-1,r+1}_{k+1,s-1}
-\frac{1}{s}E^{l,r}_{k+1,s-1}
,
\end{align}
where
\begin{align}
{F_{A}}^{l,r}_{k,s}
=&
\frac{1}{r!s!}\left(\begin{matrix}p \\ l\end{matrix}\right)
\left(\begin{matrix}p-l \\ k\end{matrix}\right)
\partial_{C^{(r-1)}}\partial_{\bar{D}^{(s)}}\partial_{[C'}C^{0}_{\cdots]}
\nn\\
&
\times
\{\textrm{Tr}_{S}\}
\left[
(D_{\mu}Y^{A})^{(l-1)}Y^{A'}(D_{\nu}Y^{\dagger}_{B})^{(k)}Y^{C(r-1)}D_{\mu'}Y^{C'}Y^{\dagger(s)}_{D}
\right]
,
\nn\\
{F_{B}}^{l,r}_{k,s}
=&
\frac{1}{r!s!}
\left(\begin{matrix}p \\ l\end{matrix}\right)
\left(\begin{matrix}p-l \\ k\end{matrix}\right)
\partial_{C^{(r)}}\partial_{\bar{D}^{(s-1)}}\partial_{[\bar{D}'}C^{0}_{\cdots]}
\nn\\
&
\times
\{\textrm{Tr}_{S}\}
\left[
(D_{\mu}Y^{A})^{(l-1)}Y^{A'}(D_{\nu}Y^{\dagger}_{B})^{(k)}Y^{C(r)}Y^{\dagger(s-1)}_{D}D_{\mu'}Y^{\dagger}_{D'}
\right]
.
\end{align}
Since the $F_{A}$- and $F_{B}$-terms depend on $dC^{0}$ but not on $C^{0}$, they are invariant under the gauge transformation \eqref{C36}.
Here we notice that the expressions $A^{l,r}_{k,s}$ and $B^{l,r}_{k,s}$ in \eqref{ABkslr} are obtained from the integration by parts using the derivation operator in $D_{\mu}Y^{A}$, therefore, such integration by parts does not reduce the number of $DY^{\dagger}$.
However, $A^{l,r}_{k,s}$ in \eqref{ABkslr} contains the expression $B^{l+1,r-1}_{k-1,s+1}$ with reduced number of $DY^{\dagger}$, hence, one should be careful in using the expression $A^{l,r}_{k,s}$ in \eqref{ABkslr}.

In analyzing $b^{l,r}_{k,s}$ with $l\ne0$ in \eqref{bkslr2}, we treat the two cases $k=0$ and $k\ne0$, separately.
For the case of $k=0$, we use the expression $A^{l,r}_{k,s}$ in \eqref{ABkslr} without $B^{l+1,r-1}_{k-1,s+1}$.
Then the following recursion relation is obtained,
\begin{align}\label{bkslr3}
b^{l,q-l}_{0,q}
=&
\frac{l}{q}G^{l,q-l}_{0,q}
-\frac{(q-l)(p+1)}{q}{F_{A}}^{l,q-l}_{0,q}
-l(p+1){F_{B}}^{l,q-l}_{0,q}
\nn\\
&
+\frac{l+1}{q}E^{l+1,q-l-1}_{0,q}
+\frac{l}{q}E^{l,q-l}_{1,q-1}
-\frac{l}{q}E^{l,q-l}_{0,q}
+\frac{(l-1)(q-l+1)}{q}A^{l-1,q-l+1}_{1,q-1}
\nn\\
&
-\frac{q-l+1}{q}b^{l-1,q-l+1}_{1,q-1}
,
\end{align}
where we have set $r=q-l$, $s=q$.
For the case $k\ne0$, we cannot use the expression $A^{l,r}_{k,s}$ in \eqref{ABkslr} due to the term $B^{l+1,r-1}_{k-1,s+1}$ with reduced number of $DY^{\dagger}$.
Instead, plugging $B^{l,r}_{k,s}$ from \eqref{ABkslr} into \eqref{bkslr2}, we obtain the other recursion relation,
\begin{align}\label{bkslr4}
b^{l,q-l}_{k,q-k}
=&
G^{l,q-l}_{k,q-k}
-\frac{(p+1)(q-k)}{k+1}{F_{B}}^{l,q-l}_{k,q-k}
\nn\\
&
-E^{l,q-l}_{k,q-k}
+E^{l,q-l}_{k+1,q-k-1}
-(q-l)A^{l,q-l}_{k,q-k}
+\frac{(l-1)(q-l+1)}{l}A^{l-1,q-l+1}_{k+1,q-k-1}
\nn\\
&
-\frac{q-l+1}{l}b^{l-1,q-l+1}_{k+1,q-k-1}
,
\end{align}
where we have set $r=q-l$, $s=q-k$.
Since the expressions in \eqref{bkslr3} and \eqref{bkslr4} cannot cover the case of $l=k=0$, we have to consider this case separately.
In this case there appear only $Y$ and $Y^{\dagger}$ originated from the Taylor expansion of $C(Y,Y^{\dagger})$.
When we set $r=s=q$, we obtain
\begin{align}\label{b0s0r}
b^{0,q}_{0,q}
=&
\frac{1}{(q!)^2}
\left(\begin{matrix}p \\ 0\end{matrix}\right)
\left(\begin{matrix}p \\ 0\end{matrix}\right)
\partial_{C^{(q)}}\partial_{\bar{D}^{(q)}}C^{0}_{\rho^{(p)}}
\{\textrm{Tr}_{S}\}
\left(
Y^{C(q)}Y^{\dagger(q)}_{D}
\right)
\nn\\
=&
\frac{1}{(q!)^2}
\left(\begin{matrix}p \\ 0\end{matrix}\right)
\left(\begin{matrix}p \\ 0\end{matrix}\right)
\partial_{C^{(q-1)}}\partial_{\bar{D}^{(q)}}\partial_{C'}C^{0}_{\rho^{(p)}}
\{\textrm{Tr}_{S}\}
\left(
Y^{C(q-1)}Y^{C'}Y^{\dagger(q)}_{D}
\right)
\nn\\
=&
(p+1)F^{0,q}_{0,q}
+\frac{1}{q}E^{1,q-1}_{0,q}
.
\end{align}
Here we have defined a gauge invariant quantity,
\begin{align}
F^{0,q}_{0,q}
=
\frac{1}{(q!)^2}
\left(\begin{matrix}p \\ 0\end{matrix}\right)
\left(\begin{matrix}p \\ 0\end{matrix}\right)
\partial_{C^{(q-1)}}\partial_{\bar{D}^{(q)}}\partial_{[C'}C^{0}_{\rho^{(p)}]}
\{\textrm{Tr}_{S}\}
\left(
Y^{C(q-1)}Y^{C'}Y^{\dagger(q)}_{D}
\right)
\end{align}
through the anti-symmetrization of the $p$-form gauge field,
\begin{align}
\partial_{C'}C^{0}_{\rho^{(p)}}
=
(p+1)\partial_{[C'}C^{0}_{\rho^{(p)}]}
+\partial_{\rho_{1}}C^{0}_{C'\rho_{2}\cdots}
+\cdots
.
\end{align}

Now, for a fixed value of $q$ and $m$ in \eqref{Sp3}, the recursion relations \eqref{bkslr3} and \eqref{bkslr4} lead to
\begin{align}\label{bsum}
\sum_{k=0}^{m}b^{m-k,q-m+k}_{k,q-k}
=&
\sum_{k=0}^{m}\frac{m-k}{q}G^{m-k,q-m+k}_{k,q-k}
\nn\\
&
+\sum_{k=0}^{m}\frac{m-k+1}{q}E^{m-k+1,q-m+k-1}_{k,q-k}
-\sum_{k=0}^{m}\frac{m-k}{q}E^{m-k,q-m+k}_{k,q-k}
.
\end{align}
We omitted the dependence of $F_{A}$- and $F_{B}$-terms in \eqref{bsum} since they are generically gauge invariant under the Abelian gauge transformation.
We notice that \eqref{bsum} does not involve the $A$-terms because those terms are nicely canceled out between \eqref{bkslr3} and \eqref{bkslr4}.
The relation \eqref{bsum} is still valid for the $m=0$ case since one can exactly reproduce the relation \eqref{b0s0r} by setting $m=0$ in \eqref{bsum}.
The expression \eqref{bsum} is not gauge invariant due to the presence of $E$-terms.
However, summing over all possible $m$ we find that the dependence of $E$-terms does cancel out.
Eventually, we obtain the following gauge invariant relation for a fixed $q$,
\begin{align}\label{msum}
\sum_{m=0}^{p}\sum_{k=0}^{m}b^{m-k,q-m+k}_{k,q-k}
=
\sum_{m=0}^{p}\sum_{k=0}^{m}\frac{m-k}{q}G^{m-k,q-m+k}_{k,q-k}
,
\end{align}
by omitting $F_{A}$- and $F_{B}$-terms.
Since the $G$-terms vanish in the case of $F_{\mu\nu}=\hat{F}_{\mu\nu}=0$, inserting \eqref{msum} into \eqref{Sp3}, proves the Abelian gauge invariance of our WZ-type coupling.

We considered the pullback of $C_{p}$ to the worldvolume of the ABJM theory for our calculational convenience.
However, we haven't consider the interior product of the $p$-form gauge field with the complex scalar fields $Y$ and $Y^{\dagger}$, which are needed to couple gauge fields with rank higher than $p+1$ to multiple $p$-dimensional-branes.
In M-theory, we need such interior product to write the WZ-type coupling of $6$-form gauge field to multiple M2-branes.
The absence of such interior products in our analysis in this section implies that our results are applicable only to the $3$-form gauge field  WZ-type coupling in \eqref{3action}.
Obviously, the the Abelian gauge invariance of \eqref{3action} follows from \eqref{msum} by setting $p=3$.

\section{Abelian Gauge Invariance: $F_{\mu\nu}\ne0$ $\&$ $\hat{F}_{\mu\nu}\ne0$ Case}

In the previous section, we showed that the WZ-type coupling \eqref{Sp2} with vanishing gauge field strengths is invariant under the Abelian gauge transformation \eqref{C36}.
Once the non-Abelian gauge field strengths are turned on, i.e. $F_{\mu\nu}\ne0$ $\&$ $\hat{F}_{\mu\nu}\ne0$, the $G$-terms in \eqref{msum}, which are apparently not invariant under the Abelian gauge transformation, are non-vanishing.
Therefore, for the construction of gauge invariant WZ-type coupling, one has to deform the WZ-type coupling in \eqref{Sp2} to cancel out the gauge non-invariant piece, specifically the $G$-terms in \eqref{msum}.
To that end, we start by rewriting the $G$-term as
\begin{align}\label{Gkslr2}
G^{l,r}_{k,s}
=
-\frac{i}{2}H^{l,r}_{k,s}
+\frac{i}{2}\hat{H}^{l,r}_{k,s}
\end{align}
with
\begin{align}
H^{l,r}_{k,s}
=&
(l-1)U^{l,r}_{k,s}
\{\textrm{Tr}_{S}\}
\left[
(D_{\mu}Y^{A})^{(l-2)}
\left(
F_{\mu'\mu''}Y^{A''}
\right)
Y^{A'}(D_{\nu}Y^{\dagger}_{B})^{(k)}Y^{C(r)}Y^{\dagger(s)}_{D}
\right]
\nn\\
&
-kU^{l,r}_{k,s}
\{\textrm{Tr}_{S}\}
\left[
(D_{\mu}Y^{A})^{(l-1)}Y^{A'}(D_{\nu}Y^{\dagger}_{B})^{(k-1)}
\left(
Y^{\dagger}_{B'}F_{\mu'\nu'}
\right)
Y^{C(r)}Y^{\dagger(s)}_{D}
\right]
,
\nn\\
\hat{H}^{l,r}_{k,s}
=&
(l-1)U^{l,r}_{k,s}
\{\textrm{Tr}_{S}\}
\left[
(D_{\mu}Y^{A})^{(l-2)}
\left(
Y^{A''}\hat{F}_{\mu'\mu''}
\right)
Y^{A'}(D_{\nu}Y^{\dagger}_{B})^{(k)}Y^{C(r)}Y^{\dagger(s)}_{D}
\right]
\nn\\
&
-kU^{l,r}_{k,s}
\{\textrm{Tr}_{S}\}
\left[
(D_{\mu}Y^{A})^{(l-1)}Y^{A'}(D_{\nu}Y^{\dagger}_{B})^{(k-1)}
\left(
\hat{F}_{\mu'\nu'}Y^{\dagger}_{B'}
\right)
Y^{C(r)}Y^{\dagger(s)}_{D}
\right]
.
\end{align}
In this section, we also follow the notation for compact indexing explained in the previous section.
Since the field strengths, $F_{\mu\nu}$ and $\hat{F}_{\mu\nu}$, appear in a symmetric way, we only deal with $H^{l,r}_{k,s}$, for simplicity. We rewrite $H^{l,r}_{k,s}$ by using the property of $\{\textrm{Tr}_{S}\}$ as 
\begin{align}\label{Hkslr}
H^{l,r}_{k,s}
=
\frac{1}{q}
\left[
k(l-1)J^{l,r}_{k,s}
-kK^{l,r}_{k,s}
+s(l-1)M^{l,r}_{k,s}
-krN^{l,r}_{k,s}
\right]
,
\end{align}
where
\begin{align}\label{JKMN}
J^{l,r}_{k,s}
=&
U^{l,r}_{k,s}
\{\textrm{Tr}_{S}\}
\left[
(D_{\mu}Y^{A})^{(l-2)}Y^{A''}(D_{\nu}Y^{\dagger}_{B})^{(k-1)}Y^{C(r)}Y^{\dagger(s)}_{D}D_{\mu''}
\left(
Y^{\dagger}_{B'}F_{\mu'\nu'}Y^{A'}
\right)
\right]
,
\nn\\
K^{l,r}_{k,s}
=&
U^{l,r}_{k,s}
\{\textrm{Tr}_{S}\}
\left[
(D_{\mu}Y^{A})^{(l-1)}(D_{\nu}Y^{\dagger}_{B})^{(k-1)}Y^{C(r)}Y^{\dagger(s)}_{D}
\left(
Y^{\dagger}_{B'}F_{\mu'\nu'}Y^{A'}
\right)
\right]
,
\nn\\
M^{l,r}_{k,s}
=&
U^{l,r}_{k,s}
\{\textrm{Tr}_{S}\}
\left[
(D_{\mu}Y^{A})^{(l-2)}Y^{A'}(D_{\nu}Y^{\dagger}_{B})^{(k)}Y^{C(r)}Y^{\dagger(s-1)}_{D}
\left(
Y^{\dagger}_{D'}F_{\mu'\mu''}Y^{A''}
\right)
\right]
,
\nn\\
N^{l,r}_{k,s}
=&
U^{l,r}_{k,s}
\{\textrm{Tr}_{S}\}
\left[
(D_{\mu}Y^{A})^{(l-1)}Y^{A'}(D_{\nu}Y^{\dagger}_{B})^{(k-1)}Y^{C(r-1)}Y^{\dagger(s)}_{D}
\left(
Y^{\dagger}_{B'}F_{\mu'\nu'}Y^{C'}
\right)
\right]
.
\end{align}
We would like to note that $J^{l,r}_{k,s}$ and $K^{l,r}_{k,s}$ contain $Y^{\dagger}FY$-terms with indices ($A'$, $B'$) which were the indices of ($DY$, $DY^{\dagger}$) before integration by parts.
For this reason, the indices ($A'$, $B'$) are contracted with the indices of the form fields $C^{0}$ in the representation of $U^{l,r}_{k,s}$ defined in \eqref{Ukslr}.
On the other hand, the indices ($C'$, $D'$) in the expression of $N^{l,r}_{k,s}$ and $M^{l,r}_{k,s}$ in \eqref{JKMN} are the indices of ($Y$, $Y^{\dagger}$) in the Taylor expansion \eqref{Taylor}.
Therefore, those indices are contracted with the indices of partial derivatives $\partial$ and $\bar{\partial}$ in the representation of $U^{l,r}_{k,s}$.
Subsequently we rewrite $M^{l,r}_{k,s}$ and $N^{l,r}_{k,s}$ in terms of $Y^{\dagger}_{B'}FY^{A'}$ through anti-symmetrization,
\begin{align}\label{antiC}
\partial_{\bar{D}}C^{0}_{A^{(l-2)}A''A'\bar{B}^{(k)}\rho^{(p-l-k)}}
=&
(p+1)\partial_{[\bar{D}}C^{0}_{A^{(l-2)}A''A'\bar{B}^{(k)}\rho^{(p-l-k)}]}
\nn\\
&
+(l-2)
\left.
\partial_{\bar{D}}C^{0}_{A^{(l-2)}A''A'\bar{B}^{(k)}\rho^{(p-l-k)}}
\right|_{\bar{D}\leftrightarrow A}
+\cdots
,
\end{align}
where $\partial_{\alpha}C^{0}_{...\beta ...}\big|_{\alpha\leftrightarrow\beta}\equiv\partial_{\beta}C^{0}_{...\alpha ...}$.
To be specific the anti-symmetrization \eqref{antiC} leads to
\begin{align}\label{Mkslr}
M^{l,r}_{k,s}
&=
(p+1){F_{M}}^{l,r}_{k,s}
-\frac{(k+1)(r+1)}{ls}K^{l-1,r+1}_{k+1,s-1}
+\frac{(k+1)(r+1)}{ls}N^{l-1,r+1}_{k+1,s-1}
\nn\\
&~~
+kQ^{l,r}_{k,s}
+\frac{(l-2)(k+1)(r+1)}{ls}R^{l-1,r+1}_{k+1,s-1}
+\frac{k+1}{s}S^{l,r}_{k+1,s-1}
,
\end{align}
where
\begin{align}
{F_{M}}^{l,r}_{k,s}
=&
\frac{1}{r!s!}
\left(\begin{matrix}p \\ l\end{matrix}\right)
\left(\begin{matrix}p-l \\ k\end{matrix}\right)
\partial_{C^{(r)}}\partial_{\bar{D}^{(s-1)}}\partial_{[\bar{D}'}C^{0}_{A^{(l-2)}A''A'\bar{B}^{(k)}\rho^{(p-l-k)}]}
\nn\\
&
\times
\{\textrm{Tr}_{S}\}
\left[
(D_{\mu}Y^{A})^{(l-2)}Y^{A'}(D_{\nu}Y^{\dagger}_{B})^{(k)}Y^{C(r)}Y^{\dagger(s-1)}_{D}
\left(
Y^{\dagger}_{D'}F_{\mu'\mu''}Y^{A''}
\right)
\right]
,
\nn\\
Q^{l,r}_{k,s}
=&
U^{l,r}_{k,s}
\{\textrm{Tr}_{S}\}
\left[
(D_{\mu}Y^{A})^{(l-2)}Y^{A''}(D_{\nu}Y^{\dagger}_{B})^{(k-1)}Y^{C(r)}Y^{\dagger(s-1)}_{D}D_{\mu''}Y^{\dagger}_{D'}
\left(
Y^{\dagger}_{B'}F_{\mu'\nu'}Y^{A'}
\right)
\right]
,
\nn\\
R^{l-1,r+1}_{k+1,s-1}
=&
U^{l-1,r+1}_{k+1,s-1}
\{\textrm{Tr}_{S}\}
\left[
(D_{\mu}Y^{A})^{(l-3)}Y^{A''}(D_{\nu}Y^{\dagger}_{B})^{(k)}Y^{C(r)}D_{\mu''}Y^{C'}Y^{\dagger(s-1)}_{D}
\left(
Y^{\dagger}_{B'}F_{\mu'\nu'}Y^{A'}
\right)
\right]
,
\nn\\
S^{l,r}_{k+1,s-1}
=&
\partial_{\mu''}U^{l,r}_{k+1,s-1}
\{\textrm{Tr}_{S}\}
\left[
(D_{\mu}Y^{A})^{(l-2)}Y^{A''}(D_{\nu}Y^{\dagger}_{B})^{(k)}Y^{C(r)}Y^{\dagger(s-1)}_{D}
\left(
Y^{\dagger}_{B'}F_{\mu'\nu'}Y^{A'}
\right)
\right]
.
\end{align}
Manifestly, the first term in the right-hand side of \eqref{Mkslr} is invariant under the gauge transformation \eqref{C36}.
Now inserting \eqref{Mkslr} into \eqref{Hkslr} and integrating the $K$-term by parts with the help of the derivation operator in $DY$, we rewrite $H^{l,r}_{k,s}$ as
\begin{align}\label{Hkslr2}
H^{l,r}_{k,s}
=&
\frac{1}{q}
\left[
s(l-1)(p+1){F_{M}}^{l,r}_{k,s}
-k(l-1)(l-2)T^{l,r}_{k,s}
-k(l-1)(k-1)V^{l,r}_{k,s}
\right.
\nn\\
&
\quad
-klK^{l,r}_{k,s}
-\frac{(l-1)(k+1)(r+1)}{l}K^{l-1,r+1}_{k+1,s-1}
\nn\\
&
\quad
-krN^{l,r}_{k,s}
+\frac{(l-1)(k+1)(r+1)}{l}N^{l-1,r+1}_{k+1,s-1}
\nn\\
&
\quad
-rk(l-1)R^{l,r}_{k,s}
+\frac{(l-1)(l-2)(k+1)(r+1)}{l}R^{l-1,r+1}_{k+1,s-1}
\nn\\
&
\quad
\left.
-k(l-1)S^{l,r}_{k,s}
+(l-1)(k+1)S^{l,r}_{k+1,s-1}
\right],
\end{align}
where
\begin{align}
T^{l,r}_{k,s}
=&
U^{l,r}_{k,s}
\{\textrm{Tr}_{S}\}
\left[
(D_{\mu}Y^{A})^{(l-3)}D_{\mu'}D_{\mu'''}Y^{A'''}Y^{A'}(D_{\nu}Y^{\dagger}_{B})^{(k-1)}Y^{C(r)}Y^{\dagger(s)}_{D}
\left(
Y^{\dagger}_{B'}F_{\mu''\nu'}Y^{A''}
\right)
\right]
,
\nn\\
V^{l,r}_{k,s}
=&
U^{l,r}_{k,s}
\{\textrm{Tr}_{S}\}
\left[
(D_{\mu}Y^{A})^{(l-2)}Y^{A'}(D_{\nu}Y^{\dagger}_{B})^{(k-2)}D_{\mu'}D_{\nu''}Y^{\dagger}_{B''}Y^{C(r)}Y^{\dagger(s)}_{D}
\left(
Y^{\dagger}_{B'}F_{\mu''\nu'}Y^{A''}
\right)
\right]
.
\end{align}

Using the relation \eqref{Hkslr2} and following the procedure to the result in \eqref{bsum}, we obtain
\begin{align}\label{Hkslr3}
\sum_{m=0}^{p}\sum_{k=0}^{m}\frac{m-k}{q}H^{m-k,q-m+k}_{k,q-k}
=&
(p+1)\sum_{m=2}^{p}\sum_{k=0}^{m-2}\frac{(q-k)(m-k)(m-k-1)}{q^{2}}{F_{M}}^{m-k,q-m+k}_{k,q-k}
\nn\\
&
-\sum_{m=4}^{p}\sum_{k=1}^{m-3}\frac{k(m-k)(m-k-1)(m-k-2)}{q^{2}}T^{m-k,q-m+k}_{k,q-k}
\nn\\
&
-\sum_{m=4}^{p}\sum_{k=2}^{m-2}\frac{k(k-1)(m-k)(m-k-1)}{q^{2}}V^{m-k,q-m+k}_{k,q-k}
\nn\\
&
-\sum_{m=2}^{p}\sum_{k=1}^{m-1}\frac{k(m-k)}{q}K^{m-k,q-m+k}_{k,q-k}
.
\end{align}
It turns out that the $N$-, $R$-, and $S$-terms in \eqref{Hkslr2} disappear when the summation is taken over all possible $k$ and $m$ in \eqref{Hkslr3}.
For the $p=3$ case of our consideration in this paper, the $T$- and $V$-terms in \eqref{Hkslr3} do not appear since the number of worldvolume indices cannot exceed three.
Taking into account this observation and plugging \eqref{Gkslr2} and \eqref{Hkslr3} into \eqref{msum}, we obtain
\begin{align}\label{Kkslr3}
\sum_{m=0}^{3}\sum_{k=0}^{m}b^{m-k,q-m+k}_{k,q-k}
=&
\frac{i}{2}\sum_{m=2}^{3}\sum_{k=1}^{m-1}\frac{k(m-k)}{q}
\left(
K^{m-k,q-m+k}_{k,q-k}
-\hat{K}^{m-k,q-m+k}_{k,q-k}
\right)
\nn\\
&
+(\textrm{gauge\, invariant\, terms})
,
\end{align}
where $\hat{K}^{l,r}_{k,s}$ is defined as
\begin{align}
\hat{K}^{l,r}_{k,s}
=
U^{l,r}_{k,s}
\{\textrm{Tr}_{S}\}
\left[
(D_{\mu}Y^{A})^{(l-1)}(D_{\nu}Y^{\dagger}_{B})^{(k-1)}Y^{C(r)}Y^{\dagger(s)}_{D}
\left(
Y^{A'}\hat{F}_{\mu'\nu'}Y^{\dagger}_{B'}
\right)
\right]
.
\end{align}
Since the $K$-terms in \eqref{Kkslr3} are not invariant under the Abelian gauge transformation, the WZ-type coupling \eqref{Sp} is also not gauge invariant.
Therefore, in order to make the WZ-type coupling gauge invariant, we have to subtract the $K$-terms in \eqref{Kkslr3} from the action in \eqref{Sp}.

The forms of $K^{l,r}_{k,s}$ and $\hat{K}^{l,r}_{k,s}$ are obtained after carrying out the pullback \eqref{CpPB} and the Taylor expansion \eqref{Taylor}.
We want to find a compact expression of these terms before the pullback and the Taylor expansion.
To do that, we rewrite the $K^{l,r}_{k,s}$ as
\begin{align}\label{Kkslr4}
K^{l,r}_{k,s}
=&
\frac{1}{r!s!}
\left(\begin{matrix}p \\ l\end{matrix}\right)
\left(\begin{matrix}p-l \\ k\end{matrix}\right)
\partial_{C^{(r)}}\partial_{\bar{D}^{(s)}}C^{0}_{A^{(l-1)}\bar{B}^{(k-1)}(A'\bar{B}')\rho^{(p-l-k)}}
\nn\\
&
\times
\{\textrm{Tr}_{S}\}
\left[
(D_{\mu}Y^{A})^{(l-1)}(D_{\nu}Y^{\dagger}_{B})^{(k-1)}Y^{C(r)}Y^{\dagger(s)}_{D}
\left(
Y^{\dagger}_{B'}F_{\mu'\nu'}Y^{A'}
\right)
\right]
,
\end{align}
where we have replaced $m-k$ by $l$ in \eqref{Kkslr3}.
Then using the relation
\begin{align}
kl \left(\begin{matrix}p \\ l\end{matrix}\right)
\left(\begin{matrix}p-l \\ k\end{matrix}\right)
=
2\,\left(\begin{matrix}p \\ 2\end{matrix}\right)
\left(\begin{matrix}p-2 \\ l-1\end{matrix}\right)
\left(\begin{matrix}(p-2)-(l-1) \\ k-1\end{matrix}\right),
\end{align}
we rewrite the gauge non-invariant quantity in \eqref{Kkslr3} as
\begin{align}\label{Kkslr5}
&
\sum_{q=1}^{\infty}\sum_{m=2}^{3}\sum_{k=1}^{m-1}\frac{k(m-k)\lambda^{2(q+1)}}{q}K^{m-k,q-m+k}_{k,q-k}
\nn\\
&=
2\sum_{q=0}^{\infty}\sum_{m=0}^{1}\sum_{k=0}^{m}
\bigg\{
\frac{\lambda^{2(q+1)}}{q+1}\frac{1}{(q-m+k)!(q-k)!}
\,\left(\begin{matrix}p \\ 2\end{matrix}\right)
\left(\begin{matrix}p-2 \\ m-k\end{matrix}\right)
\left(\begin{matrix}(p-2)-m+k \\ k\end{matrix}\right)
\nn\\
&\hskip 2.8cm
\times\partial_{C^{(q-m+k)}}\partial_{\bar{D}^{(q-k)}}C^{0}_{A^{(m-k)}\bar{B}^{(k)}(A'\bar{B}')\rho^{(p-2-m)}}
\nn\\
&\hskip 2.8cm
\times
\{\textrm{Tr}_{S}\}
\left[
(D_{\mu}Y^{A})^{(m-k)}(D_{\nu}Y^{\dagger}_{B})^{(k)}Y^{C(q-m+k)}Y^{\dagger(q-k)}_{D}
\left(
Y^{\dagger}_{B'}F_{\mu'\nu'}Y^{A'}
\right)
\right]
\bigg\}
.
\end{align}
The expression \eqref{Kkslr5} involves the scalar fields $Y$ and $Y^{\dagger}$ originated from the Taylor expansion ($Y^{C}$, $Y^{\dagger}_{D}$) and the pullback ($Y^{A'}$, $Y^{\dagger}_{B'}$) of form fields. 
Since the worldvolume field strengths, $F_{\mu\nu}$ and $\hat{F}_{\mu\nu}$, appear via the integration by parts of covariant derivatives, they can only couple with the scalar fields from the pullback of form fields.
Keeping in mind this observation, we have the relation,
\begin{align}\label{(q+1)}
&
\frac{1}{q+1}
\{\textrm{Tr}_{S}\}
\left[
(D_{\mu}Y^{A})^{(m-k)}(D_{\nu}Y^{\dagger}_{B})^{(k)}Y^{C(q-m+k)}Y^{\dagger(q-k)}_{D}
\left(
Y^{\dagger}_{B'}F_{\mu'\nu'}Y^{A'}
\right)
\right]
\nn\\
&~~~=
\{\textrm{Tr}_{S}\}
\left[
(D_{\mu}Y^{A})^{(m-k)}(D_{\nu}Y^{\dagger}_{B})^{(k)}Y^{C(q-m+k)}Y^{\dagger(q-k)}_{D}
\left(
Y^{\dagger}_{B'}Y^{A'}
\right)
F_{\mu'\nu'}
\right]
.
\end{align}
Substitution of \eqref{(q+1)} into \eqref{Kkslr5} gives
\begin{align}\label{Kkslr6}
&
\sum_{q=1}^{\infty}\sum_{m=2}^{3}\sum_{k=1}^{m-1}\frac{k(m-k)\lambda^{2(q+1)}}{q}K^{m-k,q-m+k}_{k,q-k}
\nn\\
&=
2\sum_{m=0}^{1}\sum_{k=0}^{m}\lambda^{l+k+2}
\left(\begin{matrix}p \\ 2\end{matrix}\right)
\left(\begin{matrix}p-2 \\ m-k\end{matrix}\right)
\left(\begin{matrix}p-2-m+k \\ k\end{matrix}\right)
\nn \\
&\hskip 2.2cm\times\{\textrm{Tr}_{S}\}
\left[
\textrm{i}_{Y}\textrm{i}_{Y^{\dagger}}C_{A^{(m-k)}\bar{B}^{(k)}\rho^{(p-m)}}(D_{\mu}Y^{A})^{(m-k)}(D_{\nu}Y^{\dagger}_{B})^{(k)}F_{\mu'\nu'}
\right]
\nn\\
&=
2\lambda^{2}
\left(\begin{matrix}p \\ 2\end{matrix}\right)
\{\textrm{Tr}_{S}\}
\left(
P[\textrm{i}_{Y}\textrm{i}_{Y^{\dagger}}C_{\rho^{(p)}}]F_{\mu'\nu'}
\right)
\nn\\
&=
2
\{\textrm{Tr}_{S}\}
\left(
P[\lambda^{2}\textrm{i}_{Y}\textrm{i}_{Y^{\dagger}}C_{(p)}]_{(p-2)}\wedge F
\right)
,
\end{align}
where we introduce an interior product for a $p$-form field 
$\Omega^{(p)}$, 
\begin{align}
\textrm{i}_{Y}\textrm{i}_{Y^{\dagger}}\Omega^{(p)}
=
\textrm{i}_{Y}\Omega^{(p)}_{\cdots\bar{B}'}Y^{\dagger}_{B'}
=
\Omega^{(p)}_{\cdots A'\bar{B}'}Y^{A'}Y^{\dagger}_{B'}
=
-\textrm{i}_{Y^{\dagger}}\textrm{i}_{Y}\Omega^{(p)}
.
\end{align}

From \eqref{Sp3}, \eqref{Kkslr3}, and \eqref{Kkslr6}, we read the counter term to cancel out the gauge dependent piece in a compact form with $p=3$,
\begin{align}\label{CTterm}
S_{\textrm{c.t.}}
=
-\mu_{2}\int_{3}
\{\textrm{Tr}_{S}\}
[P[i\lambda^{2}(\textrm{i}_{Y}\textrm{i}_{Y^{\dagger}})C_{(3)}]
\wedge(F-\hat{F})].
\end{align}
Here explicit expressions including $\{ {\rm Tr}_S\}$ in \eqref{CTterm} are given by
\begin{align}
\{{\rm Tr}_S\}[C_{\mu A\bar B}Y^{A}Y_{B}^{\dagger}F_{\nu\rho}]
=&
C_{\mu A\bar{B}}{}^{\hat{a}bd}_{a\hat{b}c}
Y^A{}^{a}{}_{\hat{a}}
Y_{B}^{\dagger}{}^{\hat{b}}{}_{b}
F_{\mu\nu}{}^{c}{}_{d},
\nn\\
\{{\rm Tr}_S\}[C_{\mu A\bar{B}}Y^{A}Y_{B}^{\dagger}\hat{F}_{\nu\rho}]
=&
C_{\mu A\bar{B}}{}^{\hat{a}b\hat{d}}_{a\hat{b}\hat{c}}
Y^A{}^{a}{}_{\hat{a}}
Y_{B}^{\dagger}{}^{\hat{b}}{}_{b}
\hat{F}_{\mu\nu}{}^{\hat{c}}{}_{\hat{d}}.
\end{align}
For more details of 
$C_{\mu A\bar{B}}{}^{\hat{a}bd}_{a\hat{b}c}$ and 
$C_{\mu A\bar{B}}{}^{\hat{a}b\hat{d}}_{a\hat{b}\hat{c}}$, 
see (2.7) in \cite{Kim:2010hj}. 
Similar form of counter term with \eqref{CTterm} 
was also obtained in \cite{Lambert:2009qw},  
in which the form fields are not functionals of scalar fields. 
Addition of the counter term \eqref{CTterm} to the action \eqref{Sp} finally defines the gauge invariant WZ-type coupling for the 3-form gauge field,
\begin{align}\label{Sp6}
S_{3}
=
\frac{\mu_{2}}{2}\int_{3}
\{\textrm{Tr}_{S}\}
\left[
P[C_{(3)}]
+(\textrm{c.c.})
\right]
-\frac{\mu_{2}}{2}\int_{3}
\{\textrm{Tr}_{S}\}
\left[
P[i\lambda^{2}(\textrm{i}_{Y}\textrm{i}_{Y^{\dagger}})C_{(3)}]\wedge(F-\hat{F})
+(\textrm{c.c.})
\right]
.
\end{align}

\section{Conclusion}

This paper is a complement of the program started in \cite{Kim:2010hj}.
The objective of the program is to construct the WZ-type couplings describing the dynamics of multiple M2-branes in non-trivial 3- and 6-form fields in 11-dimensional supergravity.

In \cite{Kim:2010hj} we constructed the WZ-type couplings preserving the U$_{\textrm{L}}(N)\times$U$_{\textrm{R}}(N)$ non-Abelian gauge symmetry of the ABJM theory.
This was achieved by appropriately choosing the scalar field dependence of the form fields and selecting single traces from all possible contractions of non-Abelian gauge indices.
After circle compactification, these restrictions successfully reproduced the Myers couplings with symmetrized-trace in type IIA string theory.

The WZ-type couplings should preserve not only the non-Abelian gauge symmetries of the worldvolume theory but also the Abelian gauge symmetries of the bulk 11-dimensional supergravity.
The action should be invariant under {\it the Abelian gauge transformations} \eqref{C36} of the 3- and the 6-form gauge fields.
The verification of this invariance is what was missing in \cite{Kim:2010hj}.

In this paper, we concentrate on the WZ-type coupling for the 3-form gauge field and showed that the WZ-type coupling in \cite{Kim:2010hj} is invariant under the Abelian gauge transformation only when the non-Abelian gauge field strengths vanish.
In the case of non-vanishing non-Abelian field strengths, we identified a modification by the terms involving those field strengths, in order to make the WZ-type coupling invariant under the Abelian gauge transformation.
We also found that the constructed gauge invariant 3-form coupling is expressed in a compact form \eqref{Sp6}.
Extension of our study in this paper to the cases of the 6-form gauge field and the non-linear form fields would be interesting.

\section*{Acknowledgements}

This work was supported by the Korea Research Foundation Grant funded by the Korean Government with grant numbers NRF-2014R1A1A2057066 (Y.K.), NRF-2014R1A1A2059761 (O.K.) and the Mid-career Researcher Program through the NRF grant funded by the Korean government (MEST) (No. 2014-051185) (O.K.), the grant MIUR 2010YJ2NYW\!\_\!\_001 (D.D.T).

\end{document}